\documentclass{aastex6}

\usepackage{color}

\def\msol{\,{\rm M}_\odot}              
\def\rad{r_\mathrm{d,AD}}   

\usepackage{epstopdf}
\begin{document}

\title{Magnetically self-regulated formation of early protoplanetary discs}

\author
{Patrick Hennebelle }
\affil{Laboratoire AIM, 
Paris-Saclay, CEA/IRFU/SAp - CNRS - Universit\'e Paris Diderot, 91191, 
Gif-sur-Yvette Cedex, France \\
Laboratoire de radioastronomie, UMR CNRS 8112,\\
 \'Ecole normale sup\'erieure et Observatoire de Paris,
24 rue Lhomond, 75231 Paris cedex 05, France }

\author{Beno\^it Commer{\c c}on}
\affil{\'Ecole normale sup\'erieure de Lyon,
CRAL, UMR CNRS 5574,\\ Universit\'e de Lyon, 69364 Lyon Cedex 07,  France}

\author{Gilles Chabrier}
\affil{\'Ecole normale sup\'erieure de Lyon,
CRAL, UMR CNRS 5574,\\ Universit\'e de Lyon, 69364 Lyon Cedex 07,  France, \\
School of Physics, University of Exeter, Exeter, EX4 4QL, UK }

\and

\author{Pierre Marchand}
\affil{\'Ecole normale sup\'erieure de Lyon,
CRAL, UMR CNRS 5574,\\ Universit\'e de Lyon, 69364 Lyon Cedex 07,  France}


\date{}


\begin{abstract}
The formation of protoplanetary discs  during the collapse 
of molecular dense cores is significantly influenced by
angular momentum transport, notably by the magnetic torque.
In turn, the evolution of the magnetic field is determined by dynamical processes and non-ideal MHD effects 
such as ambipolar diffusion.
Considering simple relations between various timescales characteristic of the magnetized collapse,
we derive an expression for the early disc radius,  
$ r  \simeq  18 \, {\rm AU}  \;   \left( { \eta_{\rm AD} / 0.1 \, {\rm s}   } \right)^{2/9} 
\left( {B_z / 0.1\, {\rm G}} \right) ^{-4/9}  \left( { M / 0.1 \msol} \right) ^{1/3},$
where $M$ is the total disc plus protostar mass,  $\eta_\mathrm{AD}$ is the ambipolar diffusion coefficient
and $B_z$ is the magnetic field in the inner part of the core. This is about significantly smaller than the  
discs that would form if angular momentum was conserved. 
 The analytical predictions are confronted against 
a large sample of 3D, non-ideal MHD collapse calculations covering variations of a factor 100 in core mass, a factor 10 in the level of turbulence, a factor 5 in rotation, and magnetic mass-to-flux over critical mass-to-flux ratios 2 and 5.
The disc radius estimates are found to agree with the numerical simulations within less than a factor 2.
 A striking prediction of our analysis is the weak dependence of circumstellar disc radii upon the
various relevant quantities, suggesting weak variations among class-0 disc sizes.
In some cases, we note the onset of large spiral arms beyond this radius.

\end{abstract}


\section{Introduction} \label{sec:intro}

Circumstellar discs are of fundamental importance in astrophysics because 
they are the birth sites of planet formation. Yet, our current understanding of centrifugally 
supported discs lack a clear description of how and when they 
form. The exact role played by magnetic field, in particular, remains an unsetlled issue. Various teams have been consistently 
finding that catastrophic braking may occur when the magnetic field and the rotation 
axis are aligned \citep{Allen2003,Galli2006,Mellon2008,Hennebelle2008,Price2007}.
In such circumstances, magnetic braking can be so intense that the formation 
of primordial discs at the class-0 stage can be suppressed even for modest magnetizations. 
Although recent observations have revealed that large discs are  rare, if not absent, 
at class-0 stage \citep{Maury2010,tobin2015}
complete inhibition of disc formation cannot be considered as a plausible scenario because of (i) the assumed aligned 
configuration \citep{Hennebelle2009,Joos2012,Joos2013,Santos-Lima2012,Seifried2012} and (ii) the ideal MHD assumption 
\citep{Dapp2010,Krasnopolsky2012,Li2011,Tomida2015,Tsukamoto2015,Wurster2016,masson2016}. Indeed, when either misalignment, turbulence or non-ideal 
MHD effects are included, discs tend to form more easily, although by no means as large as in pure hydrodynamical calculations \citep{Tomida2015,masson2016}.

Naively, a broad distribution of disc properties might be expected, depending for instance on the core mass, the
amount of rotation or turbulence in the core, the strength of the field and its configuration. 
In this paper, we derive a theoretical framework which suggests the opposite, i.e. that discs at their early stages are remarkably regulated by a combination 
of magnetic braking and non-ideal MHD effects, leading to similar sizes. In \S\ref{flux_distrib} and~\ref{radius_sec}, we develop
simple analytical arguments leading to our suggestion.
In \S\ref{simu_sec}, we compare these analytical estimates of the disc sizes to a series of collapse calculations corresponding to a large 
variety of initial conditions.
Section \ref{conclusion} concludes the paper.

\section{Magnetic flux distribution}
\label{flux_distrib}

We first aim at assessing the intensity of magnetic braking. Recent 3D simulations 
\citep{Tomida2015,masson2016} found that ambipolar diffusion leads to a 
plateau, i.e. a nearly uniform magnetisation, in the inner part of the collapsing cores, 
with typical values of the order of 0.1 G  for 1 $\msol$ cores, up to $\sim$ 0.3 G or so for 100 $\msol$ cores. To understand this property we 
assume stationarity, and reduce the problem to one radia dimension   (all quantities in the following are simply written $x\equiv x(r)$). 
In that case, the Faraday equation reduces to 
\begin{eqnarray}
v_r B_z \simeq { c^2 \eta_{\rm AD} \partial_r B_z \over 4 \pi}, 
\label{eq_bz}  
\end{eqnarray}
where $v_r$ denotes the radial velocity, $B_z$ the field vertical (poloidal) component and $\eta_{\rm AD}$ the ambipolar diffusivity.

Let us remind that  in ideal MHD, flux and mass conservation inside concentric cylinders lead to $B_z \propto \Sigma \simeq 2 \rho h$, where $\Sigma$
is the column density and $h$ the typical thickness. Assuming mechanical equilibrium, we get $h \simeq C_{\rm s} / \sqrt{\pi G \rho}$ and therefore
$B_z \propto \sqrt{\rho}$.

\begin{figure}
\center{\includegraphics[angle=0,width=6in]{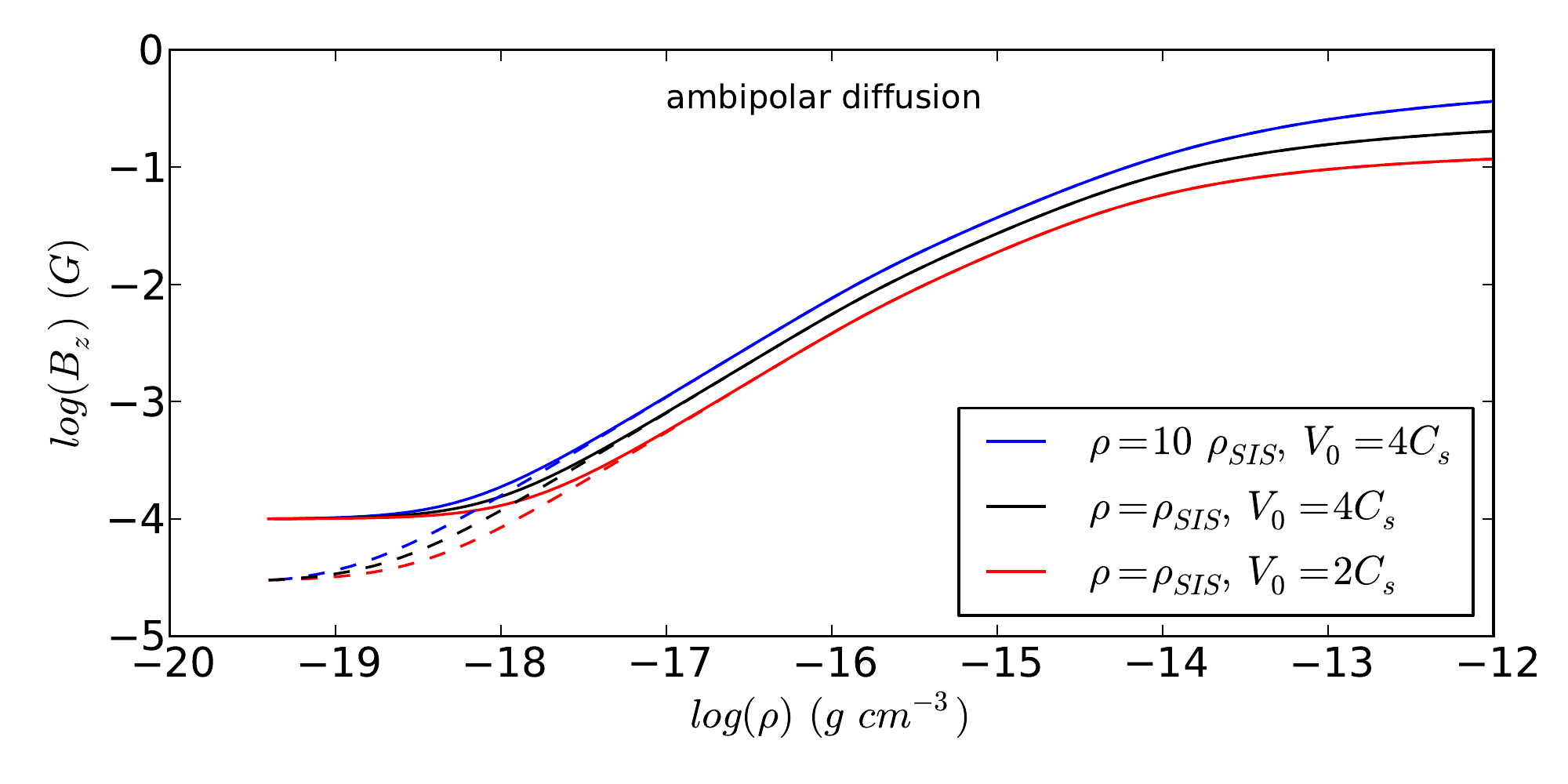}} 
\center{\includegraphics[angle=0,width=6in]{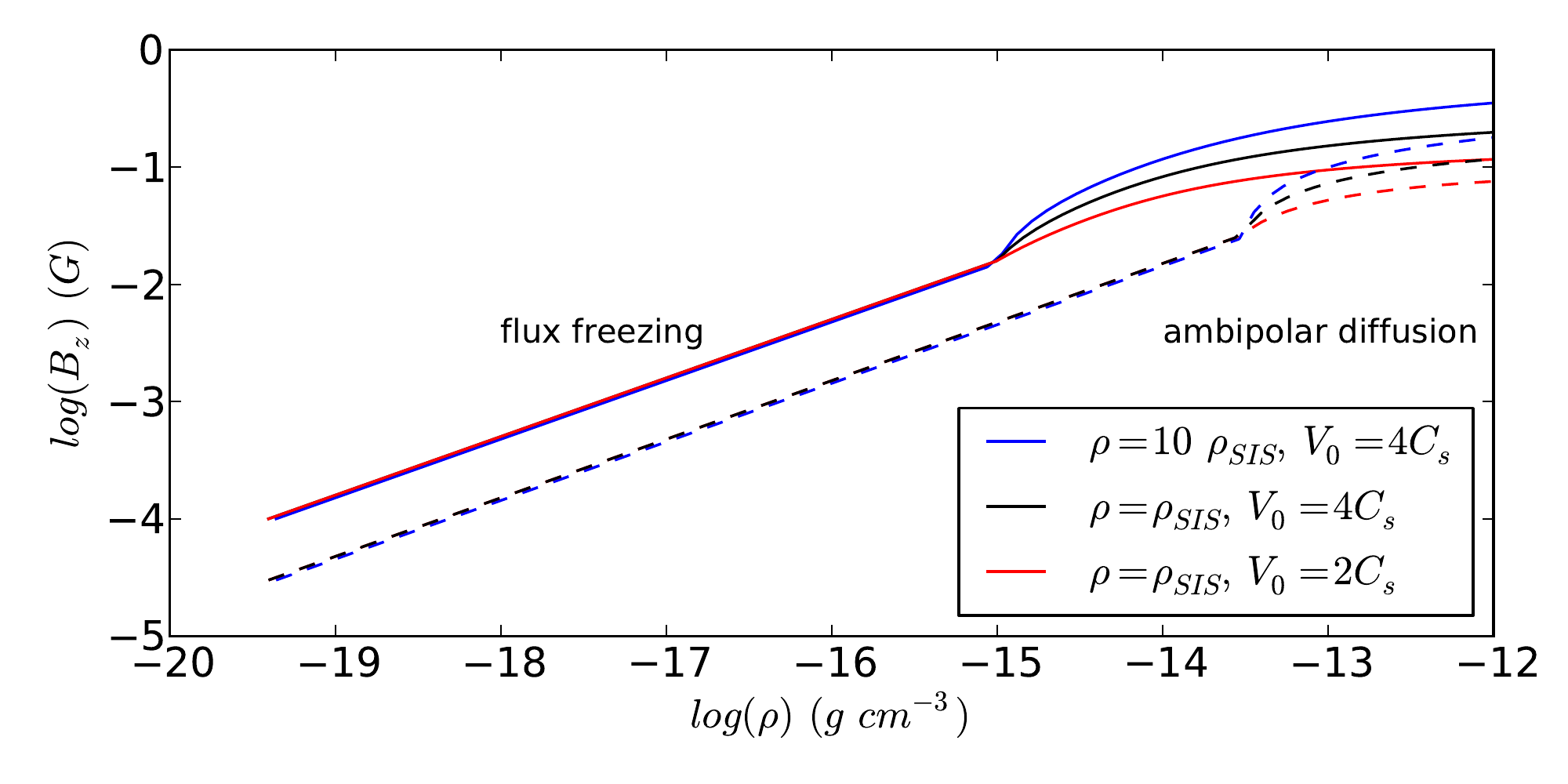}} 
\center{\includegraphics[angle=0,width=6in]{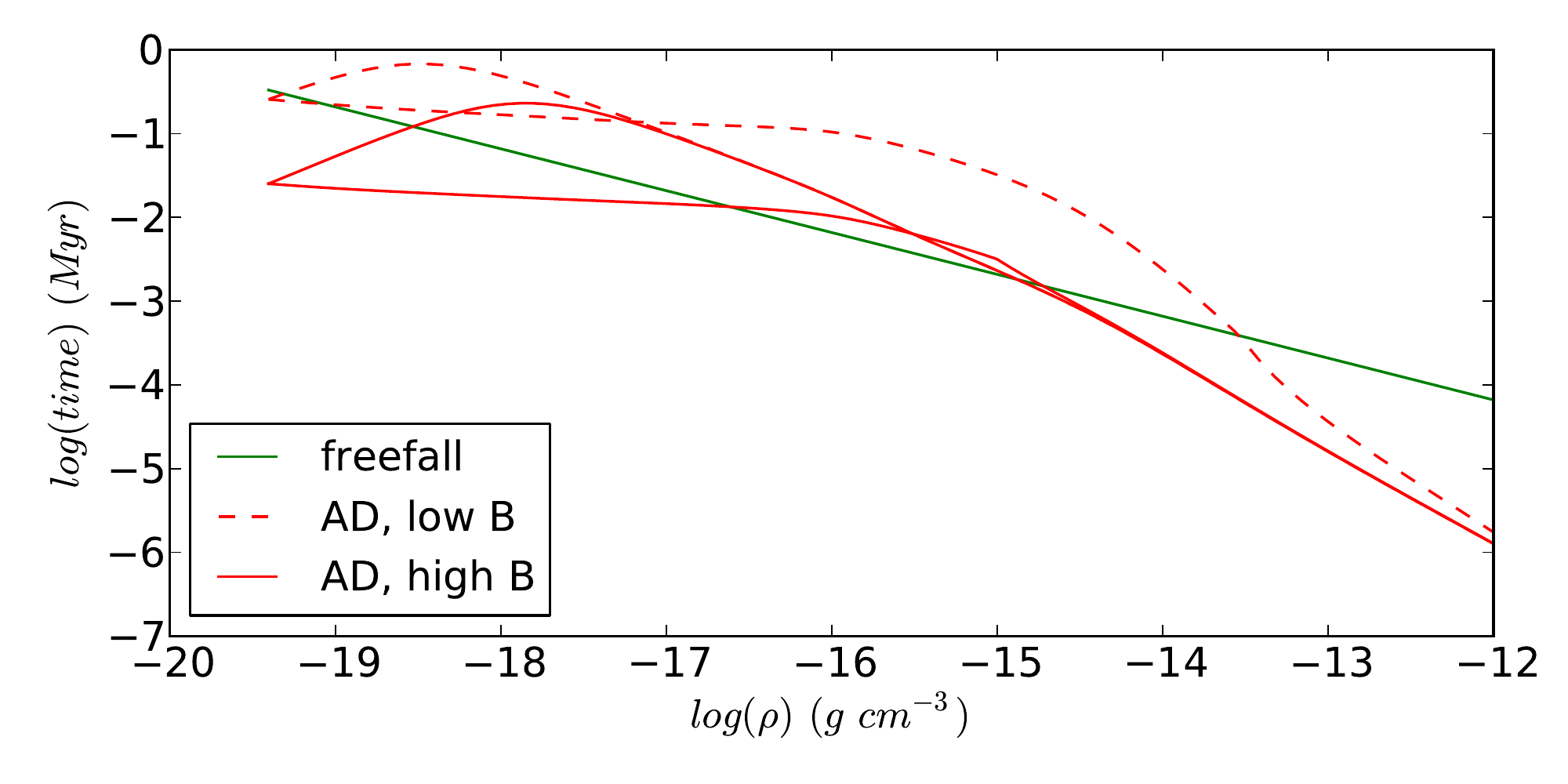}} 
\caption{
Upper panel: Vertical magnetic field component given by eqn.~\ref{eq_bz}, $B_z$, as a function of density $\rho$, for 
two magnetic intensities (solid lines: 100 $\mu G$,
dashed lines: 30 $\mu G$) at the core boundary  and three density and velocity profiles.
Middle panel: same as top one except that flux freezing is assumed until $t_\mathrm{ff} \simeq t_\mathrm{AD}$.
The value of $B_z$ varies only by a factor of a few.
Lower panel: freefall and ambipolar times for the 4 models displayed in red in the 
middle and lower panels. At high densities the freefall time is longer than the ambipolar diffusion time.  }
\label{Bz}
\end{figure}


Because of the complex dependence of the resistivity $\eta_{\rm AD}$ upon magnetic intensity $B$ and density $n$ \citep{Marchand2016}, 
eqn.~(\ref{eq_bz}) cannot be solved analytically. 
To get a  solution of eqn.~(\ref{eq_bz}), we calculated a table of resistivities for 
a series of densities and magnetic intensities from \citet{Marchand2016}, from which we get $\eta_{\rm AD}$ for any 
values by interpolation. 
To integrate numerically eqn.~(\ref{eq_bz}), the density and the radial velocity must be specified  and we set  \citep[e.g.][]{larson1969,shu1977} 
\begin{eqnarray}
\nonumber
\rho(r) = \delta {C_{\rm s}^2 \over 2 \pi G r^{2} }, \\
v_r(r) = V_0 (r/r_0)^{-1/2}. 
\label{field}
\end{eqnarray}
We considered 2 values of the external field, namely $B_z=30$ and 100 $\mu$G, as well as 3 densities 
and velocity field amplitudes, two typical of low-mass cores and one typical of high-mass ones.
 For the low-mass cores, we took $\delta=1$ (i.e. the singular isothermal sphere, sis) and 
$V_0=2 \times C_{\rm s}$ at $r_0=10$ AU and twice this values (which corresponds to a faster collapse). 
For the high-mass core, we took $\delta=10$ and $V_0=4 \times C_{\rm s}$.

Figure~\ref{Bz} displays the results.
Since in real cores, the flux distribution is due to a combination of  flux freezing and ambipolar diffusion, we explore two cases. 
First, we solve eqn.~(\ref{eq_bz}) from the edge to the center of the core (upper panel). Second, we 
assume flux freezing up to the point where the freefall, $t_\mathrm{ff}=\sqrt{3 \pi / (32 G \rho)} $, 
 and the ambipolar diffusion times, $t_\mathrm{AD}=4 \pi / (c^2 \eta_\mathrm{AD} r^2)$, become comparable ($10^{-15}$ g cm$^{-3}$
for 100 $\mu$G and  $3 \times 10^{-14}$ g cm$^{-3}$ for 30 $\mu$G see
lower panel), then we solve eqn~(\ref{eq_bz}) (middle panel). 
  Red lines correspond to $V_0=2 \times C_{\rm s}$ and dark ones to $V_0=4 \times C_{\rm s}$.
The dashed lines display the low magnetisation cases and solid lines the high magnetisation ones.
 As seen, the value of the magnetic intensity outside the core
has a weak influence on the value inside it. A clear transition occurs between a slightly sublinear regime 
(where $B_z \propto \, \sim \rho^{2/3}$) to a plateau at $\rho \simeq 10^{-15}$g cm$^{-3}$. 
From Fig.~5 of \citet{Marchand2016}, we see that indeed $\eta_{\rm AD}$ displays
 two different regimes, which correspond to densities respectively below and above 
 $\simeq 10^{-15}$ g cm$^{-3}$.
The lower panel of Fig. 1 shows the freefall time and the ambipolar diffusion time for the 4 low-mass cases displayed by the red lines in the
upper and middle panels. Clearly, while the freefall time is shorter than the ambipolar diffusion time in the outer part of the core, 
the reverse is true in the core inner part \citep{nakano2002}.
The magnetic field at the center of the core weakly depends on the physical conditions 
and remains remarkably constant. The assumption of stationarity is also well justified as the freefall time is much longer and 
the ambipolar diffusion one. 
These features 
agree quite well with the 3D simulations performed by \citet{masson2016} (their Fig.~1) and \citet{Tomida2015}. 
Furthermore, we see that a  slowly collapsing low-mass core (red lines) has a smaller central 
magnetic intensity than a more rapidly collapsing one (black lines).
The massive cores, which have both a large inward velocity and a large density, display
even higher central magnetic intensities.

Altogether the variations of the magnetic field in the inner part of the core remain limited and weakly depend 
on the initial conditions.

\section{Theoretical estimate of the disc radius}
\label{radius_sec}
To obtain an estimate of the disc radius, we examine the relevant timescales and we estimate the various quantities 
at the disc centrifugal radius location, i.e. at the disc-envelope boundary. Let us stress that the envelope and 
the disc are connected by an accretion shock that is quite thin. 
Therefore, outside the disc the gas in its vicinity is nearly in freefall (see for example 
Figs.~3 and 4 of \citet{Hennebelle2008}) and the results of Sect.~\ref{flux_distrib} can be applied. 

\subsection{Timescales and equilibria}
The first important  timescales are the ones that control the evolution of the azimuthal magnetic field, $B_\phi$, which is
 responsible for the magnetic braking. On one hand, $B_\phi$ is generated by the differential rotation on a timescale 
$\tau_{\rm far}$, and on the other hand it is diffused vertically by ambipolar diffusion on a timescale $\tau_{\rm diff}$, with
\begin{eqnarray}
\tau _{\rm far} &\simeq &{ B_\phi h \over B_z v_\phi} , \nonumber\\
\tau _{\rm diff} &\simeq &{ 4 \pi  h^2 \over c^2  \eta_{\rm AD}  } {  B_z^2 + B_\phi ^2 \over B_z^2} \simeq { 4 \pi  h^2 \over c^2  \eta_{\rm AD}  },
\label{tau1}
\end{eqnarray}
where $h$ denotes the thickness of the disc.  

The second relevant timescales are the  magnetic braking one and the rotation time. They are given by
\begin{eqnarray}
\nonumber
\tau _{\rm br} &\simeq& { \rho v_\phi 4 \pi h \over B_z B_\phi}, \\
\tau _{\rm rot} &\simeq& { 2 \pi r \over v_ \phi},
\end{eqnarray}
where $r\equiv r_{\rm d}$ denotes the disc radius.

Then, we assume that the gas in the neighbourhood of the disc outer part has 
a Keplerian velocity (in practice it may be a little lower) and is roughly in vertical hydrostatic equilibrium:
\begin{eqnarray}
v_\phi &\simeq& \sqrt{ G (M_* + M_{\rm  d}) \over r}, \\
h &\simeq& {C_{\rm s} \over \sqrt{4 \pi G (\rho + \rho_*)} }, 
\end{eqnarray}
where $M_{\rm d}$ is the mass of the disc, $M_*$ the mass of the central star and $\rho_* = M_* / (4 \pi) r_*^{-3}$.

Finally, the density in the envelope is given by
\begin{eqnarray}
\rho(r) = \delta { C_{\rm s}^2 \over 2 \pi G r^2} \left( 1 + {1 \over 2} \left( { v_\phi (r)\over C_{\rm s}} \right)^2 \right). 
\label{rho_exp}
\end{eqnarray}
Apart for $\delta$, which is a coefficient on the order of a few, the first term simply correponds to the singular isothermal sphere
\citep{shu1977} while the second one is a correction that must be included when rotation is 
significant, particularly in the inner part of the envelope close to the disc edge, as
discussed in \citet{hennebelle2004} (see their appendix and Fig.~2).
Note that  for massive stars, 
$\delta$ may be  up to about 10 as shown in Fig.~3 of \citet{hennebelle2011}.

\subsection{Dependence of the disc radius}

The disc properties are the result of the balance between various quantities at the disc-envelope boundary. 
First of all, as mentioned above, the generation of the toroidal field through 
differential rotation is  offset by the ambipolar diffusion in the vertical direction.
From eqns.(\ref{tau1}), with $\tau _\mathrm{far} \simeq \tau_\mathrm{diff}$, we  get

\begin{eqnarray}
{B_\phi \over h v_\phi} &\simeq& { 4 \pi \over c^2 \eta_\mathrm{AD} }  B_z.
\label{far}
\end{eqnarray}

Second of all,  the braking and the rotation 
timescales must be of same order, $\tau _\mathrm{br} \simeq \tau_\mathrm{rot}$, yielding
\begin{eqnarray}
{B_\phi}  \simeq {2 h  \rho  \over r } v_\phi ^2 B_z^{-1}.
\label{br}
\end{eqnarray}

\noindent Combining eqs.~(\ref{far}) and~(\ref{br}) yields
\begin{eqnarray}
{2  \rho  \over r } v_\phi  \simeq { 4 \pi \over c^2 \eta_\mathrm{AD} }  B_z^2,
\label{vdhr}
\end{eqnarray}
while vertical and radial equilibria at the disc outermost limit implies

\begin{eqnarray}
{ \delta   G ^{1/2} (M_\mathrm{d} + M_*) ^{3/2} \over 2 \pi  r^{9/2} }  \simeq { 4 \pi \over c^2 \eta_\mathrm{AD} }  B_z^2,
\end{eqnarray}
where, for sake of simplicity, we have assumed $\rho \propto v_\phi^2$ in eqn.~(\ref{rho_exp}).

All these relations lead to
\begin{eqnarray}
   \rad  \simeq   \left(  {\delta G ^{1/2} c^2 \eta_\mathrm{AD}  \over 8 \pi^2 } \right)^{2/9} 
B_z^{-4/9}  (M_\mathrm{d} + M_*)^{1/3}.
\label{rad}
\end{eqnarray}

The mass of the star/disc system, $M_\mathrm{d} + M_*$,
 grows as the envelope gets accreted. We take 0.1 $M_\odot$ as a fiducial value since we are investigating the 
class-0 phase.

\noindent With these values, eqn.~(\ref{rad}) can be rewritten:

\begin{eqnarray}
\nonumber
 \rad  \simeq  18 \, {\rm AU}  \;   & \times & \\ 
\delta ^{2/9 }\left( {   \eta_\mathrm{AD} \over 0.1 \, {\rm s}   } \right)^{2/9} &&
\left( {B_z \over 0.1\,{\rm G} } \right) ^{-4/9}   \left( { M_\mathrm{d} + M_* \over 0.1 \msol} \right) ^{1/3}.
\label{radius}
\end{eqnarray}



The striking result illustrated by eqn.~(\ref{radius}) is the weak dependence of the disc radius upon all involved quantities.
Note that, in principle, 
  the magnetic resistivity $\eta_\mathrm{AD}$ depends on  density (see Fig.~5 of \citet{Marchand2016}),
but  this dependence is very shallow. We find a more pronounced, although still moderate dependence of the radius upon $B$ as $\sim B_z^{-0.5}$.
In principle this could introduce some variations among disc radii but, as seen in Sect.~\ref{flux_distrib}, the magnetic field in the inner part of the envelope of the
cores is also regulated by ambipolar diffusion.
Finally, the radius depends also weakly on the mass.  Indeed,  as accretion proceeds, the disc is expected to become 
only about twice larger when the star becomes 10 times more massive, i.e. $M_*=1\msol$. 

We also note that $C_{\rm s}$ does not enter explicitly in eqn.~(\ref{radius}), suggesting weak dependence of the disc radius 
upon the velocity field, be it purely thermal or turbulent (through an effective sound speed 
$C_{\rm s}^{\rm eff}=(C_{\rm s}+\langle v_{\rm rms}^2\rangle^{1/2})^{1/2}$),
as indeed found in the simulations (see below). In practice, some dependence on the various supports enters in 
the coefficient $\delta$ but since it appears at the power $2/9$ this leads to weak variations. \\

It is interesting to compare these trends with the  purely hydrodynamical case. Let us consider a spherical cloud of density 
$\rho_0$ in solid body rotation at a rate $\Omega_0$.
When a fluid particle initially at radius $R_0$ reaches centrigugal equilibrium into the disc, its radius is  
\begin{eqnarray}
r_\mathrm{d,hydro} \simeq {\Omega_0^2 R_0^4 \over 4 \pi / 3 \rho_0 R_0^3 G } = 3 \beta R_0  = 106 \, {\rm AU} \, 
{\beta  \over 0.02 }
\, \left( {M \over 0.1 \msol}\right) ^{1/3}  \left( {\rho_0 \over 10^{-18} {\rm g \, cm}^{-3} }\right)^{-1/3},
\end{eqnarray}
 where $\beta =R_0^3 \Omega_0^2/ 3GM$ denotes the core rotational support. Whereas the mass dependence remains the same as above, the radius now strongly (quadratically) depends of the initial rotation rate. As cores have a typical $\beta \simeq 0.02$ 
\citep{goodman1993,belloche2013}, purely 
hydrodynamical discs should be on average significantly (typically 5-6 times larger) than the ones we predict.

\setlength{\unitlength}{1cm}
\begin{figure*} 
\begin{picture} (0,8)
\put(0,0){\includegraphics[width=8cm]{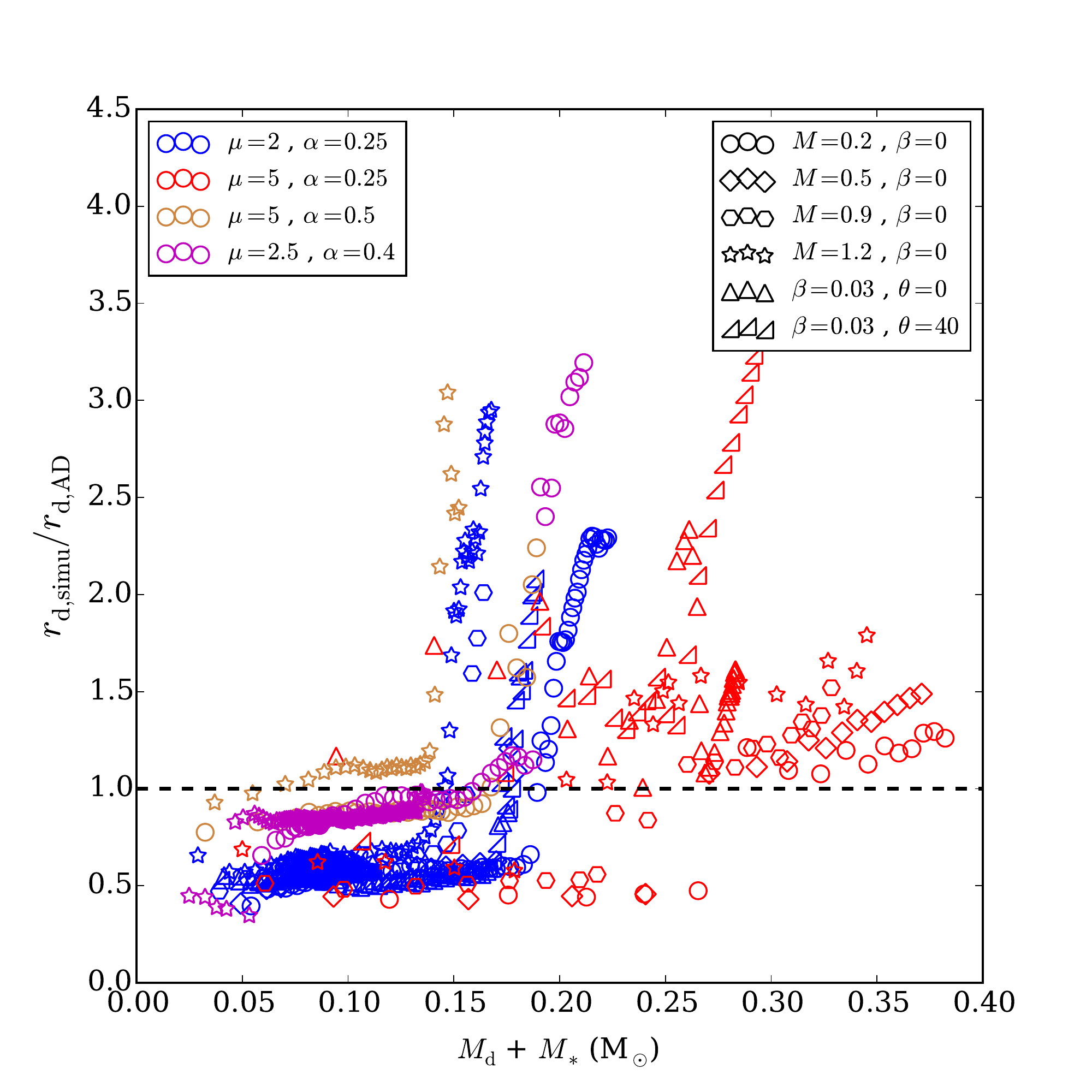}}  
\put(8,0){\includegraphics[width=8cm]{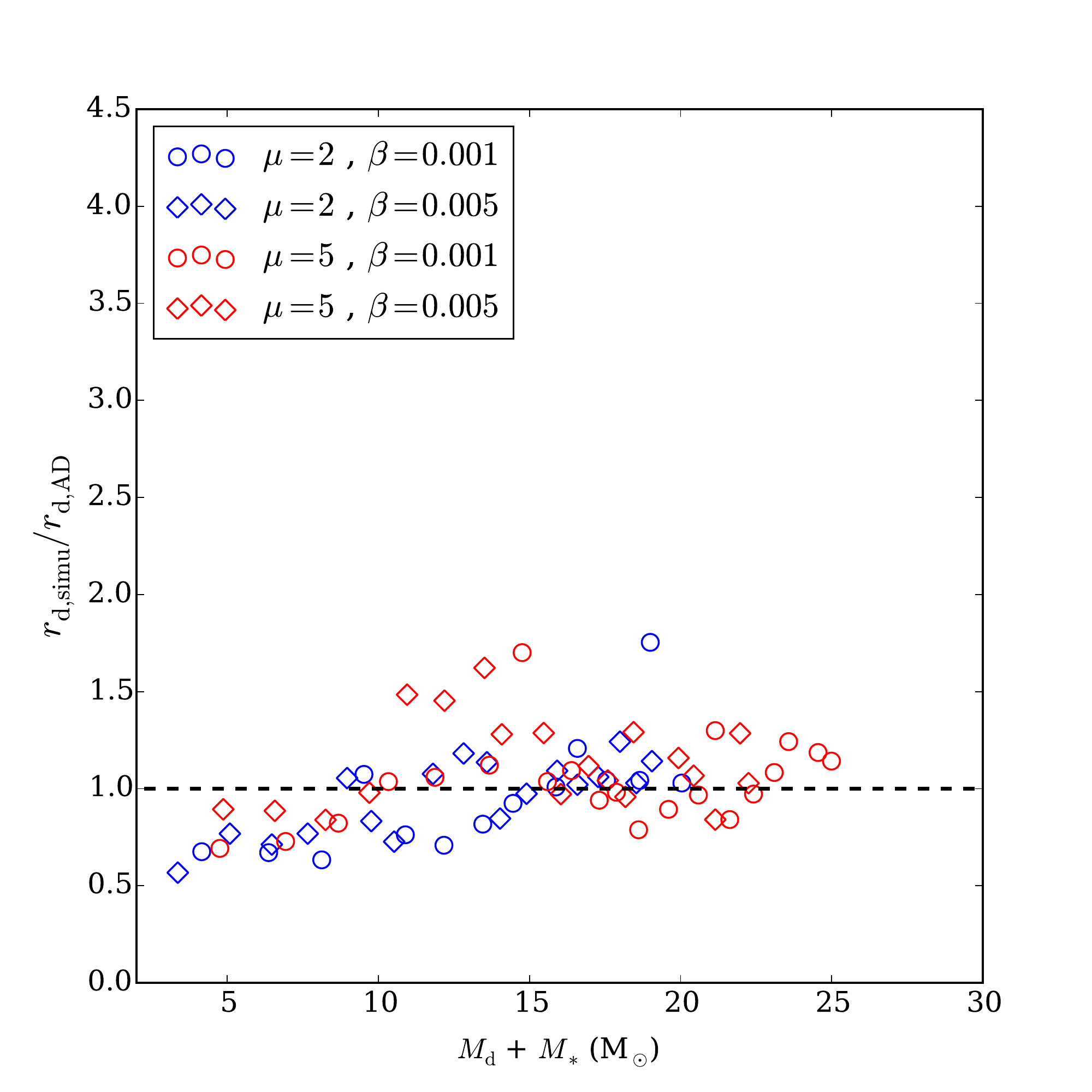}}  
\end{picture}
\caption{Ratio of the disc radius measured in the simulations over the theoretical estimate (eq.~(\ref{radius})) 
as a function of the total (disc plus star) mass. The left panel corresponds to the low mass cores and the right panel to the 
high mass ones. As seen, the ratio is on average of the order of 1. The points that deviate significantly
($r_\mathrm{d,simu}/r_\mathrm{d,AD} \simeq 2-3$) correspond to the development of prominent spiral patterns which connect
to the disc (see text).} 
\label{simu1}
\end{figure*}



\section{Comparison with numerical simulations}
\label{simu_sec}

\subsection{Initial conditions}
To test the validity of our analytical model, we have performed two series of numerical simulations of 
non-ideal MHD collapse with ambipolar diffusion, with the RAMSES code  
\citep{teyssier2002,fromang2006,masson2012}. 

The first type of simulations are identical and/or similar to the ones performed in 
\citet{masson2016}. They have an initial core mass of 1 $\msol$, a uniform density profile and a uniform magnetic field 
with a mass-to-flux over critical mass-to-flux ratio of 2 or 5.
We considered various levels of turbulence, ranging from ${\cal M}=0.2$ to ${\cal M}=1.2$,  different values of 
$\alpha$ (thermal over 
gravitational energy) and $\beta$ and different angles $\theta$ between the initial magnetic field and the rotation axis. 
For the second type of simulations, we considered a massive core of 100 $\msol$ with a uniform temperature of 20~K. 
The initial density profile  follows $\rho(r)=\rho_\mathrm{c}/(1+(r/r_\mathrm{c})^{-2})$, 
where  $\rho_\mathrm{c}\sim 7.7\times10^{-18}$~g~cm$^{-3}$and the extent of the central plateau is $r_\mathrm{c}=0.02$~pc. 
The initial core radius is $r_0=0.2$~pc. 
Radiative transfer is properly accounted for in the simulations, as in \citet{Commercon2011}, and takes into account
 the feedback from protostellar luminosity using pre-main sequence evolution models \citep{hosokawa2010}  attached to sink particles.
The coarser grid resolution is $64^3$ and we allow for 9 additional levels of refinement, which gives a minimum resolution of 5~AU (the sink accretion radius is then of 20~AU).

\subsection{Results}
The disc radius is defined according to the criteria described in \citet{Joos2012}. We first perform an azimuthal average of the rotation,
radial velocity and sound speed. We then select the rings for which both the radial velocity and the sound speed are smaller than 
50$\%$ of the rotation velocity. 
Figure~\ref{simu1}  displays the ratio of the disc radius measured  in the simulations at different times over the radius inferred from eqn. (\ref{radius}), as a function of the 
total (star+disc) system mass. 
The left panel correspond to 1 $\msol$ mass cores and the right one to 100 $\msol$ mass ones. The central mass corresponds either to the mass of the first 
Larson core in the low-mass models, or to the mass of the sink particle in the high-mass ones. For these latter simulations, we also considered
 a value of 0.3 G for $B_z$ in eqn. (\ref{radius}), as mentioned in \S2.1.
As seen in the figures, the 
agreement between the theoretical predictions and the simulation is globally quite satisfactory. Most of the points lie between 0.5 and 2 
indicating that our theoretical estimate $\rad$, given by eqn.~(\ref{radius}), agrees within less than a factor 2 with the numerical results. 

For some simulations,  notably for the low-mass cores, we see a sudden and steep increase of the radius by a factor 2-3 above some mass. 
 We verified that this occurs when the disc and stellar mass reaches about 30 to 50\% of the prestellar mass, 
depending on the various parameters,
and the estimated Toomre parameter becomes much smaller than unity.
This behaviour thus corresponds to the non-linear development 
of spiral patterns, which connect to the disc (see Fig.~8 and 11 of \citet{masson2016}), making the definition of a disc radius 
rather ambiguous. The dynamics of 
these patterns clearly differs from the one of an axi-symmetric disc.

\section{Conclusion}
\label{conclusion}

In this paper we proposed simple analytical arguments
for the formation of early circumstellar discs in collapsing magnetized cores,
 suggesting that the discs are self-regulated by the magnetic braking and the 
ambipolar diffusion. The disc radius estimates derived from the theory have been compared to the values
obtained from a series of non-ideal MHD simulations, 
covering a large range of masses, turbulent support, geometrical configurations and magnetic intensities.
The comparisons show an agreement between the theoretical and numerical results, within less than a factor 2.
The most striking result is the weak dependence of the disc size upon the core mass, the intensity of the field
or the level of turbulence in the core, suggesting small variations between class-0 disc sizes under different environments. Clearly, further observations should be able to probe this prediction.

\acknowledgments
We thank the anonymous referee for a helpful report.
This research has received funding from the European Research Council under the European
 Community's Seventh Framework Programme (FP7/2007-2013 Grant Agreement no. 247060 and no. 306483). We acknowledge financial support from "Programme National de Physique Stellaire" (PNPS) of CNRS/INSU, France.
 We thank Rolf Kuiper for providing tabulated PMS evolution tracks. 

\vspace{5mm}




\begin{thebibliography}{55}



\bibitem[{{Allen} {et~al.}(2003){Allen}, {Li}, \& {Shu}}]{Allen2003}
{Allen}, A., {Li}, Z.-Y., \& {Shu}, F. 2003, ApJ, 599, 363




\bibitem[Belloche(2013)]{belloche2013} Belloche, A.\ 2013, EAS Publications Series, 62, 25

\bibitem[{{Commer{\c c}on} {et~al.}(2011{\natexlab{a}}){Commer{\c c}on},
  {Hennebelle}, \& {Henning}}]{Commercon2011}
{Commer{\c c}on}, B., {Hennebelle}, P., \& {Henning}, T. 2011{\natexlab{a}},
  ApJL, 742, L9


\bibitem[{{Dapp} \& {Basu}(2010)}]{Dapp2010}
{Dapp}, W. \& {Basu}, S. 2010, A\&A, 521, 56



\bibitem[{{Fromang} {et~al.}(2006){Fromang}, {Hennebelle}, \&
  {Teyssier}}]{fromang2006}
{Fromang}, S., {Hennebelle}, P., \& {Teyssier}, R. 2006, A\&A, 457, 371



\bibitem[Galli et al.(2006)]{Galli2006} Galli, D., Lizano, S., Shu, F.~H., \& Allen, A.\ 2006, \apj, 647, 374 


\bibitem[Goodman et al.(1993)]{goodman1993} Goodman, A.~A., Benson, P.~J., Fuller, G.~A., \& Myers, P.~C.\ 1993, \apj, 406, 528 

\bibitem[Hennebelle et al.(2004)]{hennebelle2004} Hennebelle, P., Whitworth, A.~P., Cha, S.-H., \& Goodwin, S.~P.\ 2004, \mnras, 348, 687


\bibitem[{{Hennebelle} \& {Fromang}(2008)}]{Hennebelle2008}
{Hennebelle}, P. \& {Fromang}, S. 2008, A\&A, 477, 9



\bibitem[{{Hennebelle} \& {Ciardi}(2009)}]{Hennebelle2009}
{Hennebelle}, P. \& {Ciardi}, A. 2009, A\&A, 506, L29


\bibitem[Hennebelle et al.(2011)]{hennebelle2011} Hennebelle, P., Commer{\c c}on, B., Joos, M., et al.\ 2011, \aap, 528, A72 

\bibitem[{{Hosokawa} {et~al.}(2010){{Hosokawa}, {Yorke}, \& {Omukai}}}]{hosokawa2010}
{Hosokawa}, T., {Yorke}, H.~W., \& {Omukai}, K. 2010, \apj, 721, 478


\bibitem[{{Joos} {et~al.}(2012){Joos}, {Hennebelle}, \& {Ciardi}}]{Joos2012}
{Joos}, M., {Hennebelle}, P., \& {Ciardi}, A. 2012, A\&A, 543, 128


\bibitem[{{Joos} {et~al.}(2013){Joos}, {Hennebelle}, \& {Ciardi}}]{Joos2013}
{Joos}, M., {Hennebelle}, P.,  {Ciardi}, A., \& {Fromang}, S.  2013, A\&A, 554, 17


\bibitem[{{Krasnopolsky} {et~al.}(2012){Krasnopolsky2012}, {Li}, {Shang}, \&
  {Zhao}}]{Krasnopolsky2012}
{Krasnopolsky}, R., {Li}, Z.-Y., {Shang}, H., \& {Zhao}, B. 2012, ApJ, 757, 77


\bibitem[Larson(1969)]{larson1969} Larson, R.~B.\ 1969, \mnras, 145, 271 



\bibitem[{{Li} {et~al.}(2011){Li}, {Krasnopolsky}, \& {Shang}}]{Li2011}
{Li}, Z.-Y., {Krasnopolsky}, R., \& {Shang}, H. 2011, ApJ, 738, 180





\bibitem[Marchand et al.(2016)]{Marchand2016} Marchand, P., Masson, J., Chabrier, G., et al.\ 2016, \aap, 592, A18 

\bibitem[Masson et al.(2012)]{masson2012} Masson, J., Teyssier, R., Mulet-Marquis, C., Hennebelle, P., Chabrier, G.\ 2012, ApJS, 201, 24

\bibitem[Masson et al.(2016)]{masson2016} Masson, J., Chabrier, G., Hennebelle, P., Vaytet, N., \& Commer{\c c}on, B.\ 2016, \aap, 587, A32


\bibitem[{{Maury} {et~al.}(2010){Maury2010}, {Andr{\'e}}, {Hennebelle}, {Motte},
  {Stamatellos}, {Bate}, {Belloche}, {Duch{\^e}ne}, \& {Whitworth}}]{Maury2010}
{Maury}, A.~J., {Andr{\'e}}, P., {Hennebelle}, P., {et~al.} 2010, A\&A, 512,
  A40+

\bibitem[{{Mellon} \& {Li}(2008)}]{Mellon2008}
{Mellon}, R.~R. \& {Li}, Z. 2008, ApJ, 681, 1356


\bibitem[Nakano et al.(2002)]{nakano2002} Nakano, T., Nishi, R., \& Umebayashi, T.\ 2002, \apj, 573, 199 

\bibitem[{{Price} \& {Bate}(2007)}]{Price2007}
{Price}, D.~J. \& {Bate}, M.~R. 2007, Ap\&SS, 311, 75

\bibitem[{{Santos-Lima} {et~al.}(2012){Santos-Lima2012}, {de Gouveia Dal Pino}, \&
  {Lazarian}}]{Santos-Lima2012}
{Santos-Lima}, R., {de Gouveia Dal Pino}, E.~M., \& {Lazarian}, A. 2012, ApJ,
  747, 21

\bibitem[{{Seifried} {et~al.}(2012){Seifried}, {Banerjee}, {Pudritz}, \&
  {Klessen}}]{Seifried2012}
{Seifried}, D., {Banerjee}, R., {Pudritz}, R.~E., \& {Klessen}, R.~S. 2012,
  MNRAS, L442


\bibitem[Shu(1977)]{shu1977} Shu, F.~H.\ 1977, \apj, 214, 488 

\bibitem[{{Teyssier}(2002)}]{teyssier2002}
{Teyssier}, R. 2002, A\&A, 385, 337

\bibitem[Tobin et al.(2015)]{tobin2015} Tobin, J.~J., Looney, L.~W., Wilner, D.~J., et al.\ 2015, \apj, 805, 125 

\bibitem[{{Tomida} {et~al.}(2010){Tomida}, {Machida}, {Saigo}, {Tomisaka}, \&
  {Matsumoto}}]{Tomida10}
{Tomida}, K., {Machida}, M.~N., {Saigo}, K., {Tomisaka}, K., \& {Matsumoto}, T.
  2010, ApJL, 725, L239


\bibitem[Tomida et al.(2015)]{Tomida2015} Tomida, K., Okuzumi, S., \& Machida, M.~N.\ 2015, \apj, 801, 117


\bibitem[Tsukamoto et al.(2015)]{Tsukamoto2015} Tsukamoto, Y., Iwasaki, K., Okuzumi, S., Machida, M.~N., \& Inutsuka, S.\ 2015, \mnras, 452, 278 


\bibitem[Wurster et al.(2016)]{Wurster2016} Wurster, J., Price, D.~J., \& Bate, M.~R.\ 2016, \mnras, 457, 1037 

\end{thebibliography}




\end{document}